\documentclass[fleqn,usenatbib]{mnras}

\usepackage[T1]{fontenc}
\usepackage{ae,aecompl}

\usepackage{graphicx}	
\usepackage{amsmath}	
\usepackage{amssymb}	
\usepackage{epstopdf} 
\usepackage[para]{threeparttable} 
\usepackage{pstricks}
\usepackage{float}
\usepackage{subcaption}
\usepackage{tabularx}

\title[Physical properties of centaur (54598) Bienor from photometry]{Physical properties of centaur (54598) Bienor from photometry}

\author[Estela Fern\'andez-Valenzuela et al.]{
E. Fern\'andez-Valenzuela,$^{1}$\thanks{E-mail: estela@iaa.es (IAA)}
J. L. Ortiz$^{1}$,
R. Duffard$^{1}$,
N. Morales$^{1}$,
P. Santos-Sanz$^{1}$\\
$^{1}$Instituto de Astrof\'isica de Andaluc\'ia, CSIC, Glorieta de la Astronom\'ia s/n, Granada 18008, Spain\\}

\date{Accepted XXX. Received YYY; in original form ZZZ}

\pubyear{2015}

\begin{document}
\label{firstpage}
\pagerange{\pageref{firstpage}--\pageref{lastpage}}
\maketitle

\begin{abstract}
We present time series photometry of Bienor in four observation campaigns from 2013 to 2016 and compare them with previous observations in the literature dating back to 2000. The results show a remarkable decline in the amplitude of the rotational light curve and in the absolute magnitude. This suggests that the angle between the rotation axis and the line of sight has changed noticeably during the last 16 years as Bienor orbits the Sun. From the light curve amplitude data we are able to determine the orientation of the rotation axis of Bienor ($\beta_p=50\pm3^{\circ}$, $\lambda_p=35\pm8^{\circ}$). We are also able to constrain the $b/a$ axial ratio of a triaxial Jacobi ellipsoidal body (with semi-axis $a>b>c$). The best fit is for $b/a=0.45\pm0.05$, which corresponds to a density value of $594^{+47}_{-35}$ kg$\,$m$^{-3}$ under the usual assumption of hydrostatic equilibrium and given that Bienor's rotational period is 9.17 h. However, the absolute magnitude of Bienor at several epochs is not well reproduced. We tested several explanations such as relaxing the hydrostatic equilibrium constraint, a large North-South asymmetry in the surface albedo of Bienor or even a ring system. When a ring system of similar characteristics to those of Chariklo and Chiron is included, we can fit both the light curve amplitude and absolute magnitude. In this case the derived axial ratio is modified to $b/a=0.37\pm0.10$. The implied density is $678^{+209}_{-100}$ kg$\,$m$^{-3}$. Also the existence of a ring is consistent with the spectroscopic detection of water ice on Bienor. Nevertheless the other explanations cannot be discarded.
\end{abstract}



\section{Introduction}
\label{introduction}

Centaurs are objects with orbits located between Jupiter's and Neptune's orbits. These bodies originally came from the Trans-Neptunian Belt and were injected to the inner part of the Solar system as a result of planetary encounters, mostly with Neptune. Accordingly, centaurs are dynamically evolved objects with unstable orbits; their lifetime is around 2.7 My \citep{Horner2004a}, and most of them may become short-period comets \citep{Horner2004b,Jewitt2008}. The first centaur to be discovered was Chiron, the second largest known to date. So far, we only have evidence of the existence of about a hundred of them, compared to the thousands of Trans-Neptunian Objects (TNOs) catalogued to date, which makes them even more unique. Centaurs and TNOs are possibly the least evolved objects of the Solar System, regarding the physical properties of their materials, due to the vast distances that separate them from the Sun; they are nevertheless collisionaly evolved \citep{Campo-Bagatin2012}. Hence, centaurs yield important information about the formation of the Solar System and its outer part.

At present, the interest in centaurs has considerably increased since the discovery of orbiting material shaped in the form of rings around two of them, Chariklo \citep{Braga-Ribas2014} and Chiron \citep{Ortiz2015}. One of the proposed scenarios for the formation of rings in centaurs is the collision with other bodies of around 10 km of effective diameter during their dynamic evolution from the Trans-Neptunian Belt across to the Neptune's orbit, although there might be other possible mechanisms \citep{Pan2016,Hyodo2016}. 

Bienor is one of the largest centaurs known to date besides the two aforementioned centaurs and all 200-km sized TNOs are thought to be collisionally evolved bodies \citep{Campo-Bagatin2012}, therefore, it is plausible that Centaurs in this size range share similar collisional and dynamical histories. Hence, Bienor may be expected to display similar properties to Chariklo and Chiron, thus the special interest raised by this object. As a result, a detailed study on Bienor's rotational light curves along with its absolute magnitude has been carried out in this work. Bienor was initially designated as 2000 QC$_{243}$, and it was discovered, as its name indicates, in year 2000. Since then, observational data were published in numerous studies on colours, absolute magnitude and other photometric and spectroscopic data \cite[e.g. ][]{Delsanti2001,Ortiz2002a,Dotto2003,Romanishin2005}. However, many aspects remained to be studied.

Here we present an extensive study of this object from the photometric point of view. Observations and data reduction are detailed in Sec. \ref{sec:observations}. The results in satisfactorily reproducing the variation of the absolute magnitude at different epochs. Diverse scenarios that might overcome this issue are studied in Secs. \ref{sec:no_equilibrium}, \ref{sec:variable_albedo} and \ref{sec:ring_system}. A general discusion is presented in Sec. \ref{sec:discussion}. Sec. \ref{sec:conclusions} closes the paper with a brief summary.

\section{Observations and data reduction}
\label{sec:observations}

We carried out four observation campaigns between 2013 and 2016 using different telescopes.  A log of the observations is shown in Table \ref{tab:telescope_characteristics}. The first observation run was executed on 2013 December 6 with the 1.5 m telescope at the Sierra Nevada Observatory (OSN) in Granada, Spain, in order to obtain Bienor's absolute magnitude. We used the 2k$\times$2k CCDT150 camera, which has a field of view of 7.1'$\times$7.1' and an image scale of 0.232"pixel$^{-1}$. The images were obtained using $V$ and $R$ bands in the Bessell's filters system and in 2$\times$2 binning mode. We calibrated the observations with the Landolt PG2213+006 field, specifically with the PG2213+006a, PG2213+006b and PG2213+006c Landolt standard stars, which share similar colours with Bienor (see Tables \ref{tab:colour_Landolt} and \ref{tab:colour_Bienor}). Twelve images of the Landolt field and three of Bienor were taken altogether in each filter (we rejected one Bienor's R-band image due to blending with a star). The Landolt stars were observed at different air masses with the aim of correcting the measurements from atmospheric extinction.

\begin{table*}
\caption{Journal of observations of Bienor from different telescopes. The R and V filters are based on the Bessell system and the r\_SDSS filter is based on the Sloan Digital Sky Survey. Abbreviations are defined as follows: exposure time (T$_{\rm E}$); number of images (N) and time on target each night (T$_{\rm obj}$).}
\begin{center}
\begin{tabular}{cccccccc}
\hline
Date           & Telescope   & Filter   & Binning    & Seeing     & T$_{\rm E}$        & N       & T$_{\rm obj}$   \\
               &             &          &            & (arcsec)   &  (seconds)   &         & (hours)      \\
\hline
\hline

 2013 Dec 6  & OSN 1.5 m    &  V      & 2$\times$2 &  1.89     &   500        & 3       & 0.42   \\
 2013 Dec 6  & OSN 1.5 m    &  R      & 2$\times$2 &  2.11     &   500        & 2       & 0.28   \\
\hline
2014 Nov 18 & CAHA 1.23 m  & Clear   & 2$\times$2 & 1.72      &       300    & 21      & 1.75   \\
2014 Nov 19 & CAHA 1.23 m  & Clear   & 2$\times$2 & 1.61      &       300    & 22      & 1.83   \\
2014 Dec 18 & CAHA 1.23 m  & Clear   & 2$\times$2 & 1.84      &       300    & 26      & 2.17   \\ 
2014 Dec 27 & NOT                &  R      & 1$\times$1 & 1.08      &       250    & 63      & 4.37   \\
2014 Dec 28 & NOT                &  R      & 1$\times$1 & 0.94      &       250    & 56      & 3.89   \\
\hline
2015 Nov 5 & CAHA 1.23 m  & Clear   & 2$\times$2 & 1.44      &       250    & 41      & 2.85   \\
2015 Nov 6 & CAHA 1.23 m  & Clear   & 2$\times$2 & 1.46      &       250    & 75      & 5.21   \\
2015 Dec 13 & NOT          & r\_SDSS & 1$\times$1 & 0.74      &       400    & 10      & 1.11   \\
\hline
2016 Aug 4 & OSN 1.5 m  & R   & 2$\times$2 & 1.77      &       300    & 22      & 1.83   \\
2016 Aug 5 & OSN 1.5 m  & R   & 2$\times$2 & 1.91    &       400    & 15      & 1.67   \\
2016 Aug 6 & OSN 1.5 m  & R   & 2$\times$2 & 1.70   &       400    & 21      & 2.33   \\
2016 Aug 7 & OSN 1.5 m  & R   & 2$\times$2 & 2.51    &       400    & 18      & 2.00   \\
2016 Aug 8 & OSN 1.5 m  & R   & 2$\times$2 & 1.79    &       400    & 19      & 2.11   \\
2016 Aug 5 & OSN 1.5 m  & V   & 2$\times$2 &  1.91  &       400    & 4      & 0.45   \\
2016 Aug 7 & OSN 1.5 m  & V   & 2$\times$2 & 2.51    &       400    & 7      & 0.78  \\
2016 Aug 8 & OSN 1.5 m  & V   & 2$\times$2 & 1.79    &       400    & 4      & 0.45 \\
\hline
\end{tabular}
\end{center}
\label{tab:telescope_characteristics}
\end{table*}

\begin{table*}
\caption{Colours of the Landolt standard stars used for calibrations during campaigns 2013 and 2016.}
\begin{tabular}{cccccccc}
\hline
\hline
Campaign &Colour &  PG2213+006A   &   & PG2213+006B  &       &  PG2213+006C  & \\		
     	\hline
 2013&$V-R$  &     $0.406\pm0.003$& & $0.4270\pm0.0008$& & $0.4260\pm0.0023$ &\\
 &$R-I$   &      $0.403\pm0.005$ & & $0.4020\pm0.0015$ & &$0.4040\pm0.0068$ &\\
\hline\hline
 &  Colour    &SA23\_433 &SA23\_435    & SA23\_438  & SA23\_440   &SA23\_443 & SA23\_444 \\
 \hline
2016&$V-R$    &$0.386\pm0.003$ &$0.4690\pm0.0013$&$0.5110\pm0.0014$&$0.4930\pm0.0029$&$0.3680\pm0.0007$&$0.5500\pm0.0065$\\
   & $R-I$     &$0.3680\pm0.0013$ &$0.4760\pm0.0013$&$0.5140\pm0.0049$&$0.4640\pm0.0012$&$0.3690\pm0.0007$&$0.5140\pm0.0105$\\
\hline
\hline
\end{tabular}
\label{tab:colour_Landolt}
\end{table*}

\begin{table*}
\caption{Colours of Bienor from data published in this work and previous literature.}
\begin{threeparttable}
\begin{tabular}{cccccccc}
\hline
\hline
Colour    &Nov. 2000$^{1}$& Aug. 2001$^{2}$& 2002$^{\dag}$ $^{3}$    &Aug. 2002$^{4}$&  Oct. 2013$^{\star}$ & Aug. 2016$^{\star}$\\		
     	\hline
$V-R$  &  $0.45\pm0.04$ &  $0.44\pm0.03$&  $0.38\pm0.06$& $0.48\pm0.04$& $0.42\pm0.07$& $0.44\pm0.07$ &\\
$R-I$   &  $0.40\pm0.07$ &  $0.47\pm0.03$ &  $0.41\pm0.06$&$0.58\pm0.06$ & && \\
\hline
\hline
\end{tabular}
\begin{tablenotes}
	\item $^{1}$ \cite{Delsanti2001}, $^{2}$ \cite{Doressoundiram2002}, $^{3}$ \cite{Bauer2003}, $^{4}$ \cite{Doressoundiram2007}.\\
	\item $^{\star}$ This work.\\
	\item $^{\dag}$ \cite{Bauer2003} observed Bienor on 29$^{th}$ October 2001 and on 13$^{th}$ June 2002.
	\end{tablenotes}
	\end{threeparttable}
\label{tab:colour_Bienor}
\end{table*}

The second and third observation campaigns were executed in order to obtain different rotational light curves within an approximate interval of a year between each other. The runs of the 2014 campaign took place on November 18 and 19 and December 18, 27 and 28 with the CAHA (Centro Astron\'omico Hispano Alem\'an) 1.23 m telescope of the Calar Alto Observatory in Almer\'ia (Spain) and the 2.5 m Nordic Optical Telescope (NOT) at Roque de los Muchachos in La Palma (Spain). The instrument used at the CAHA 1.23 m telescope was the 4k$\times$4k CCD DLR-III camera. This device has a field of view of 21.5'$\times$21.5' and an image scale of 0.314"pixel$^{-1}$. No filter was used in order to obtain the largest signal to noise ratio (SNR). The images were dithered over the detector to prevent problems in the photometry associated with bad pixels or CCD defects. The instrument used at NOT was the 2k$\times$2k ALFOSC camera (Andalucia Faint Object Spectrograh and Camera), with a field of view and an image scale of 6.4'$\times$6.4' and 0.19"pixel$^{-1}$, respectively. The images were obtained using the R-band filter in the Bessell system. A total of 188 science images were taken during the whole campaign. On the other hand, the third campaign took place on 2015 November 5 and 6 and December 13 with the same telescopes and cameras used during the 2014 campaign. No filter was used in the DLR-III camera, and r\_SDSS (Sloan Digital Sky Survey) filter was used in ALFOSC. A total of 126 science images were taken during this campaign. 

The last observation campaign took place from 2016 August 4 to 8 with the 1.5 m telescope at the Sierra Nevada Observatory (OSN) in Granada (Spain) in order to obtain Bienor's absolute magnitude and rotational light curve. The CCD camera was the same as in the first campaign. The images were obtained using $V$ and $R$ bands in the Bessell's filters system and in 2$\times$2 binning mode. A total of 95 R-band and 15 V-band science images were taken during the whole campaign. We calibrated the observations with the Landolt SA23 field, specifically with the SA23\_435, SA23\_438, SA23\_443, SA23\_444, SA23\_440 and SA23\_433 Landolt standard stars, which share similar colours with Bienor (see Tables \ref{tab:colour_Landolt} and \ref{tab:colour_Bienor}). Three images of the Landolt field were taken altogether in each filter. Bienor was observed at different air masses with the aim of correcting the measurements from atmospheric extinction.

When the time spent between observations made it possible, we aimed the telescope at the same region of the sky each night in order to keep fixed the same stellar field. This is convenient as it would permit to choose the same set of references stars for all nights in the observing runs in order to minimize systematic photometric errors. At the beginning of each observation night we took bias frames and twilight sky flat-field frames to calibrate the images. We subtracted a median bias and divided by a median flat-field corresponding to each night. Specific routines written in IDL (Interactive Data Language) were developed for this task. The routines also included the code to perform the aperture photometry of all reference stars and Bienor. The procedures we followed were identical to those described in \cite{Fernandez-Valenzuela2016}.

We tried different apertures in order to maximize the signal to noise ratio (SNR) on the object for each night and to minimize the dispersion of the photometry. We also selected a radius for the sky subtraction annulus and the width of the annulus (see Table \ref{tab:photometry_parameters}).

\begin{table*}
\caption{Parameters of the photometric analysis. Abbreviations are defined as follows: aperture radius (aper.); radius of the internal annulus for the subtraction of the sky background (an.); width of the subtraction annulus (d$_{\rm an.}$) and number of reference stars (N$_{\star}$).}

\begin{threeparttable}
\begin{tabular}{ccccc}
\hline
Date                         & aper.       & an.       & d$_{\rm an.}$   & N$_{\star}$\\
                             & (pixels)    & (pixels)    & (pixels)      &            \\
\hline
\hline
2013 Dec 6 (V-Band) & 4           & 15           & 4            &  3$\dag$  \\
2013 Dec 6 (R-Band) & 4           & 10           & 4            &  3$\dag$  \\
\hline
2014 Nov 18               & 3           & 13           & 5            &  13        \\
2014 Nov 19               & 3           & 13           & 5            &  13        \\
2014 Dec 18               & 3           & 12           & 5            &  12        \\
2014 Dec 27               & 3           & 26           & 5            &  12        \\
2014 Dec 28               & 3           & 26           & 5            &  12        \\
\hline
2015 Nov 5               & 2           &  6           & 3            &  11        \\
2015 Nov 6               & 2           &  6           & 3            &  11        \\
2015 Dec 13               & 4           & 10           & 5            &  13        \\
\hline
2016 Aug 4 (R-band) & 3 & 11      & 5  & 21  \\
2016 Aug 5 (R-band)  & 3 & 11    &  5  & 21;8$\ddag$   \\
2016 Aug 5 (V-band)  & 6 &  30  &   5    & 8 $\ddag$   \\
2016 Aug 6 (R-band)  & 3 & 11   & 5  & 21   \\
2016 Aug 7 (R-band)  & 3 & 11  &  5 & 21 \\
2016 Aug 7 (V-band)  & 7 & 22  &  5  & 8 $\ddag$  \\
2016 Aug 8 (R-band)  & 3 & 11  & 5  & 21   \\
2016 Aug 8 (V-band)  &5 & 22    &  5    & 8 $\ddag$ \\
\hline
\end{tabular}
\begin{tablenotes}
	\item $\dag$Landolt standard stars: PG2213+006a, PG2213+006b and PG2213+006c.\\
	\item $\ddag$Landolt standard stars: SA23\_435, SA23\_438, SA23\_443, SA23\_444, SA23\_440, SA23\_433.
	\end{tablenotes}
	\end{threeparttable}
\label{tab:photometry_parameters}
\end{table*}

\section{Results from observations}
\label{sec:results}

\subsection{Rotational light curves from relative photometry}
\label{sec:light_curves} 

We chose the same reference stars set within each observation run. All the stars showed a good photometric behaviour. We picked out stars which presented a wide range of brightness, and that were either brighter or fainter than the object, with the aim of studying the dispersion given by the photometric data of the object with regard to similar magnitude stars. This step enabled us to assess the quality of the photometric measurement. The number of reference stars can be seen in Table \ref{tab:photometry_parameters}. From the campaigns three different light curves were obtained.

\begin{table*}
	\centering
		\caption{Photometry results for the observations from the Calar Alto, Roque de los Muchachos and Sierra Nevada Observatories. We list the Julian Date (JD, corrected from light time); the relative magnitude (Rel. mag., in mag); the error associated (Err. in mag); the topocentric ($r_{\rm H}$) and heliocentric ($\Delta$) distances (both distances expressed in au) and the solar phase angle ($\alpha$, in deg). The full table is available online.}
		\begin{tabular}{cccccc}
		\hline\hline
    $J_{\rm D}$    & Rel. Mag. &  Err.  & r$_{\rm H}$   & $\Delta$& $\alpha$  \\
				       & (mag)     & (mag)  & (au)    & (au)    & ($^{\circ}$) \\
			\hline
2456980.16349	&	-0.0065	&	0.0243	&	16.042	&	15.169	&	1.699	\\
2456980.16866	&	-0.0175	&	0.0301	&	16.042	&	15.169	&	1.699	\\
2456980.17227	&	-0.0312	&	0.0230	&	16.042	&	15.169	&	1.699	\\
2456980.17587	&	-0.0099	&	0.0176	&	16.042	&	15.169	&	1.699	\\
2456980.17948	&	-0.0705	&	0.0277	&	16.042	&	15.169	&	1.699	\\
2456980.18309	&	-0.0432	&	0.0265	&	16.042	&	15.169	&	1.700	\\
2456980.18669	&	0.0023	&	0.0217	&	16.042	&	15.169	&	1.700	\\
				\hline\hline
		\end{tabular}
	\label{tab:photometry_result}
\end{table*}

\begin{figure}
\centering
\begin{subfigure}[b]{1.\linewidth}
    \centering
    \includegraphics[width=.99\textwidth]{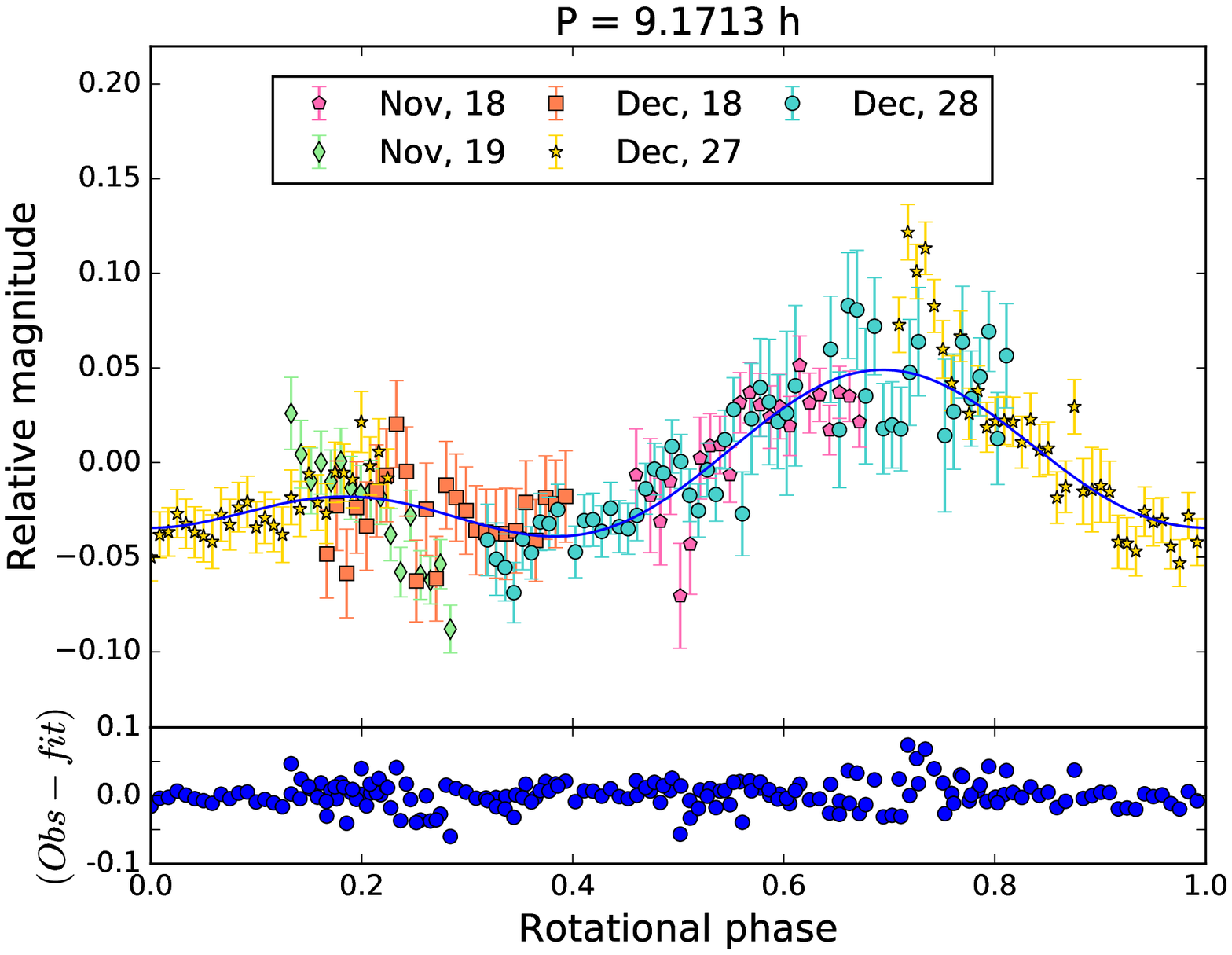}
    \caption{Rotational light curve from 18$^{th}$ and 19$^{th}$ November and from 18$^{th}$, 27$^{th}$ and 28$^{th}$ December 2014.}\label{fig:light_curve_2014}
  \end{subfigure}  \\
  \begin{subfigure}[b]{1.\linewidth}
    \centering
    \includegraphics[width=.99\textwidth]{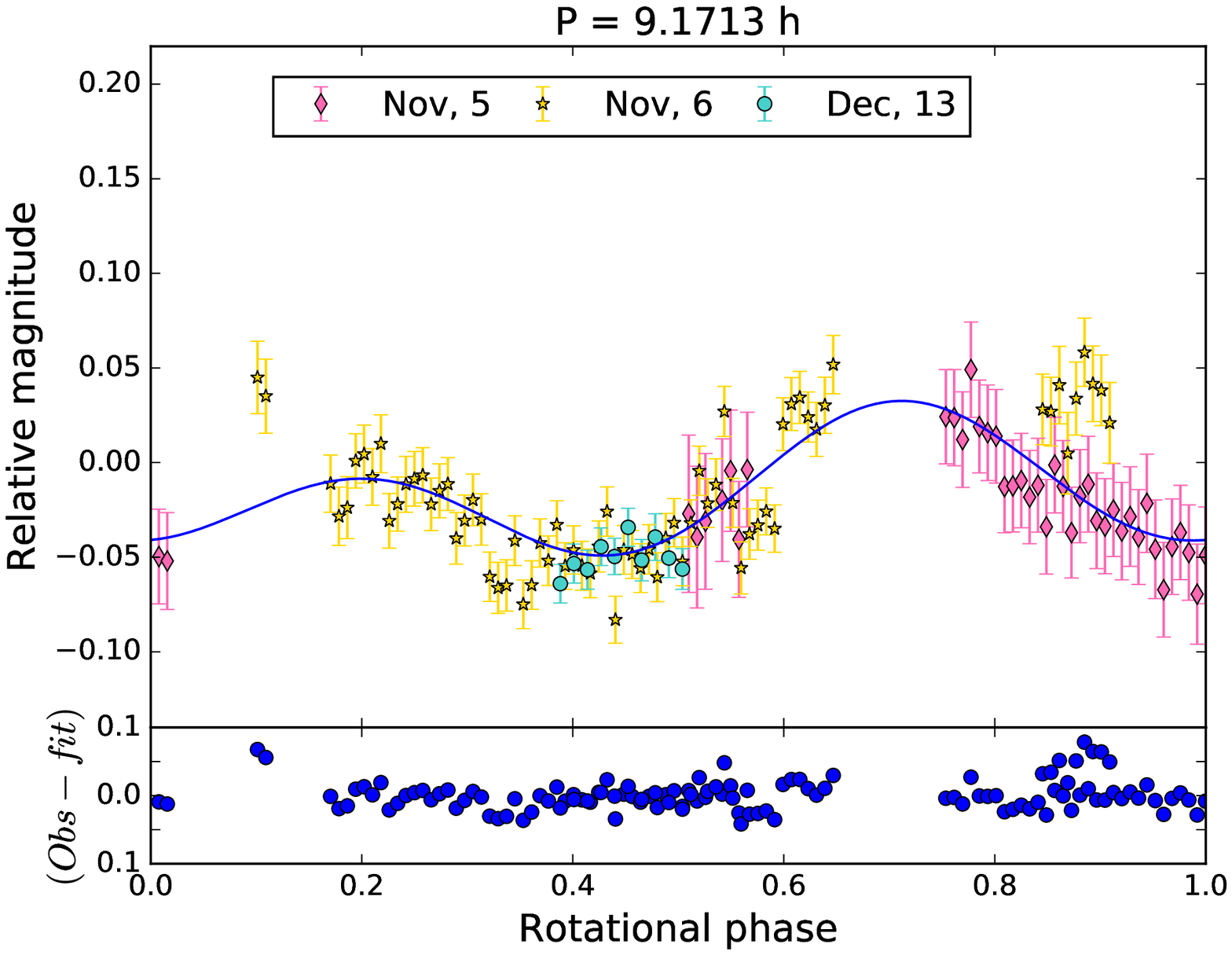}
    \caption{Rotational light curve from 5$^{th}$ and 6$^{th}$ November and 13$^{th}$ December 2015.}\label{fig:light_curve_2015}
  \end{subfigure}\\
  \begin{subfigure}[b]{1.\linewidth}
    \centering
    \includegraphics[width=.99\textwidth]{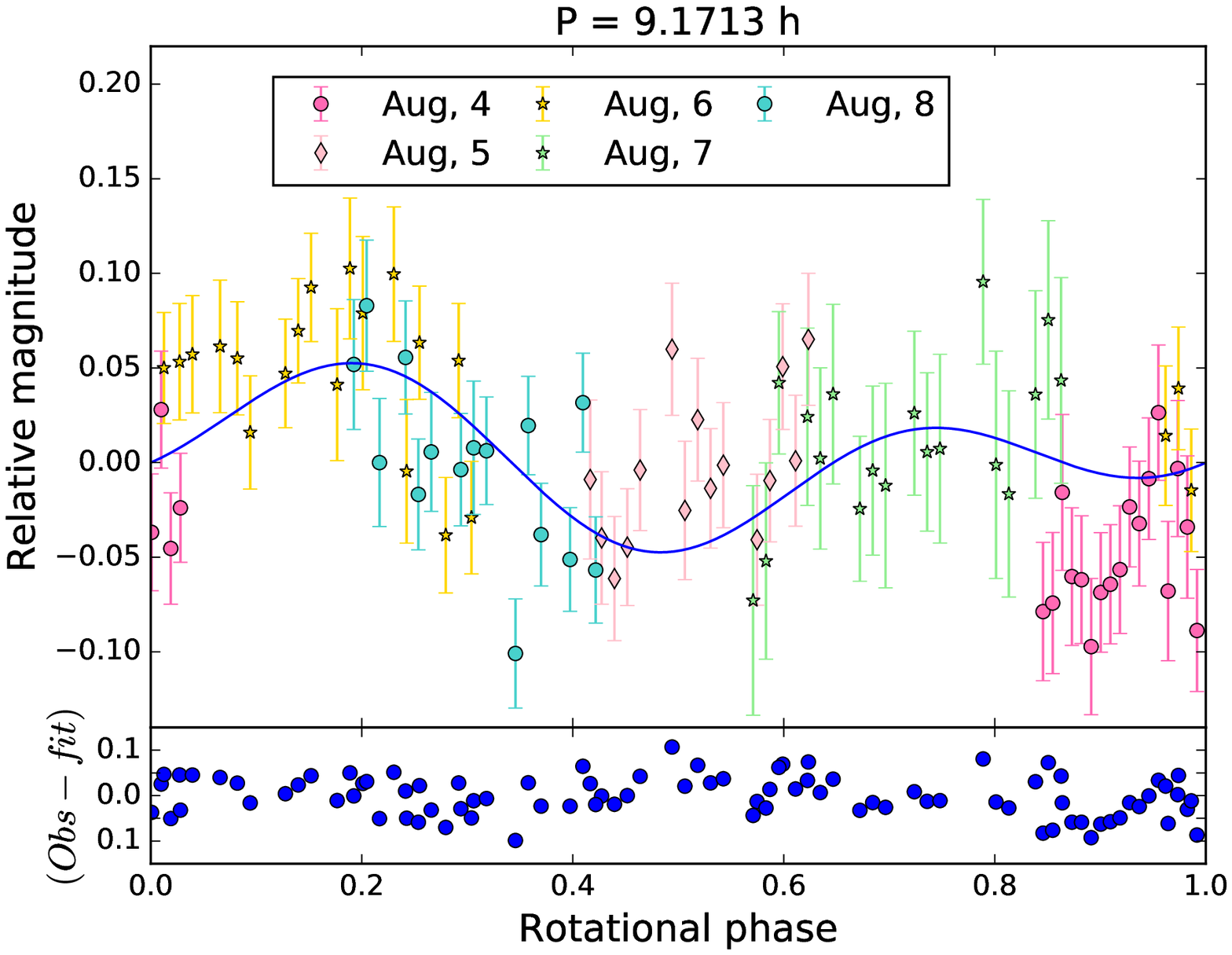}
    \caption{Rotational light curve from 4$^{th}$ to 8$^{th}$ August 2016.}\label{fig:light_curve_2016}
  \end{subfigure}%
  \caption{Rotational light curves from 2014 (upper panel), 2015 (middle panel) and 2016 (bottom panel). The points represent the observational data, each colour and symbol corresponding to a different day. The blue line shows the fit of the observational data to the second order Fourier function (Eq. \ref{eq:fourier_function}). At the bottom of each panel it can be seen the residual values of the observational data fit to the second order Fourier function (blue points).}\label{fig:rotational_light_curves}
\end{figure}

From the data we determined our own rotational period of $9.1713\pm0.0011$ h which is consistent within the error bars with that determined by \cite{Rabinowitz2007} and \cite{Ortiz2002a}. We folded the photometric data taken in 2014, 2015 and 2016 using this rotational period. In order to calculate the light-curve amplitude we fitted the data points to a second order Fourier function as follows:

\begin{equation}
\label{eq:fourier_function}
m(\phi)=a_0 + a_1\cos(2\pi\phi)+ b_1\sin(2\pi\phi)+ a_2\cos(4\pi\phi)+ b_2\cos(4\pi\phi),
\end{equation}
where $m(\phi)$ is the relative magnitude given by the fit to this equation, $\phi$ is the rotational phase and $a_0$, $a_1$, $a_2$, $b_1$ and $b_2$ are the Fourier coefficients (see Table \ref{tab:Fourier_fit_parameter}). Rotational phase is given by the following equation: $\phi=(J_{\rm D}-J_{\rm D_0})/P$; where $J_{\rm D_0}=2456980$ is an arbitrary initial Julian date corrected for light travel time, P is the target's rotational period in days and $J_{\rm D}$ is the Julian date corrected from light travel time. We obtained three light curve amplitude values: $\Delta m=0.088 \pm 0.008$ mag, $\Delta m=0.082 \pm 0.007$ and $\Delta m=0.10 \pm 0.02$ mag for the 2014, 2015 and 2016 light curves, respectively. These three light curves can be seen in Figs. \ref{fig:light_curve_2014}, \ref{fig:light_curve_2015} and \ref{fig:light_curve_2016}.

\begin{table*}
	\caption{Parameters for the second order Fourier function fits for 2014, 2015 and 2016 light curves. Columns are as follows: arbitrary initial Julian date ($J_{\rm D_0}$), second order Fourier function coefficients ($a_{0}$, $a_{1}$, $a_{2}$, $b_{1}$ and $b_{2}$) and Pearson's $\chi^2$ per degree of freedoms test ($\chi^2_{\rm PDF}$).}
	\label{tab:Fourier_fit_parameter}
	\centering
		\begin{tabular}{cccccccc}
		\hline\hline
			   Run    & $J_{\rm D_0}$  & $a_{0}$      & $a_{1}$     & $a_{2}$    & $b_{1}$     & $b_{2}$    & $\chi^2_{\rm pdf}$\\
			\hline
			   2014   &  2456980.0 & -0.00761012 &-0.00993554 &-0.03211367 &-0.01694834 & 0.0156932  &  1.55     \\
		           2015   &  2456980.0 & -0.01563793 &-0.00161743 &-0.02089169 &-0.02363507 & 0.01427225 &  1.24  \\
	 2016  &  2456980.0 &  0.00364251  & 0.02330365  & 0.01234425 & -0.02699805 & 0.01338158 &	1.77 \\           
				\hline\hline
		\end{tabular}
\end{table*}

Finally, in order to check the quality of the photometric analysis of the object, the dispersion of the residual of the fit to equation (\ref{eq:fourier_function}) was compared with the dispersion of the measurements of a star of similar flux or slightly lower than Bienor's flux (see Table \ref{tab:dispersion_results}). In order to minimize the dispersion caused by external factors, such as different CCDs or large temporal distances between observation campaigns, only those days sharing the same reference stars and also the same telescope were used for the purpose of comparing the fluxes. Therefore, only data obtained at NOT was used in the 2014 run; similarly, data obtained in November at CAHA 1.23 m were used in the 2015 run. In order to obtain the dispersion of the star, the relative magnitude was calculated with respect to the remaining reference stars. The dispersion value of the residual fit to the Fourier function is bigger in both rotational light curves in years 2014 and 2015 than the dispersion of the control star for each run. The slightly larger dispersion of Bienor's residuals compared to the dispersion of control stars may indicate a slight deficiency of the light curve modeling or intrinsic variability of Bienor at the level of $\sim$ 0.004 mag. Note that the dispersion of the data in 2016 was significantly higher than in 2014 and 2015, therefore no clear conclusion in this regard can be obtained from the 2016 data.

\begin{table}
	\caption{Comparison between the dispersion of the fit residual to the Fourier function and the dispersion of the star data with similar or lower flux than Bienor flux. Abbreviations are defined as follows: dispersion of the fit residual to the Fourier function ($\sigma_{\rm Bienor}$) and dispersion of the star residual ($\sigma_{\star}$).}
	\label{tab:dispersion_results}
	\centering
		\begin{tabular}{ccc}
		\hline\hline
                	Campaign    & $\sigma_{\rm Bienor}$  & $\sigma_{\star}$\\
			\hline
			 2014   &     0.020                         & 0.016           \\
			 2015   &     0.022                         & 0.010           \\
			 2016   &     0.050                         & 0.040            \\
     				\hline\hline
		\end{tabular}
\end{table}
\subsection{Absolute magnitude}
\label{sec:absolute_magnitude}

The absolute magnitude of a solar system body is defined as the apparent magnitude that the object would have if located at 1 AU from the Sun, 1 AU from the Earth and with 0$^{\circ}$ phase angle. This magnitude is obtained from the well-known equation:

\begin{equation}
\label{eq:absolute_magnitude}
H=m_{\rm Bienor}-5\log(r_{\rm H}\Delta)-\phi(\alpha),
\end{equation}
where $H$ is the absolute magnitude of Bienor, $m_{\rm Bienor}$ is the apparent magnitude of Bienor, $r_H$ is the heliocentric distance, $\Delta$ is the topocentric distance and $\phi(\alpha)$ is a function that depend on the phase angle. This function can be approximated by $\alpha\beta$, where $\alpha$ is the phase angle and $\beta=0.1\pm0.02$ mag$\,$deg$^{-1}$ is the phase correction coefficient, which is the average value from $\beta_V$ and $\beta_I$ given by \cite{Rabinowitz2007}. This value agrees with the value obtained in the phase angle study of \cite{Alvarez-Candal2016}. On the other hand, the apparent magnitude of Bienor is given as follows: 

\begin{equation}
\label{eq:apparent_magnitude}
m_{\rm Bienor}=m_{\rm \star_i} -\frac{5}{2}\log\left(\frac{<F_{\rm Bienor}>}{<F_{\star_i}>}\right)-k\Delta\zeta,
\end{equation}
where $m_{\star_i}$ is the apparent magnitude of Landolt standard stars (the subscript $i$ indicates different Landolt standard stars); $<F_{\rm Bienor}>$ is the average flux of Bienor; $<F_{\star_i}>$ is the average flux of different Landolt standard stars; $k$ is the extinction coefficient and $\Delta\zeta$ is the difference between the Landolt standard stars' air mass and Bienor's air mass.

We carried out a linear fit in order to obtain the extinction coefficient following the equation:

\begin{equation}
m_{\rm \star,i}=m^0_{\rm \star,i} + k\zeta,
\end{equation}
where $m_{\rm \star,i}$ is the apparent magnitude of the star and $m^0_{\rm \star,i}$ is the apparent magnitude corrected for atmospheric extinction (see Table \ref{tab:abolute_photometry}).

Finally, the values obtained for the absolute magnitudes of Bienor, during the 2013 campaign, in $V$ and $R$ band are $7.42\pm0.05$ mag and $7.00\pm0.05$ mag, respectively. On the other hand, the absolutes magnitudes, during the 2016 campaign, in $V$ and $R$ band are $7.47\pm0.04$ mag and $7.03\pm0.02$ mag. From those, we could also obtain the $(V-R)$ colour, which is $0.42\pm0.07$ mag and $0.44\pm0.07$ mag in 2013 and 2016, respectively.

\begin{table*}
	\caption{Absolute photometry results for the observations from the Sierra Nevada Observatory. We list the Julian Date ($J_{\rm D}$); the filter in the Bessell system; the absolute magnitude ($H$, in mag); the error associated to the absolute magnitude ($e_{\rm H}$,  in mag); the extinction coefficient ($k$); the error associated to the extinction coefficient ($e_{\rm k}$); the air mass ($\zeta$); the topocentric ($r_H$) and heliocentric ($\Delta$) distances (both distances expressed in au) and the solar phase angle ($\alpha$, in deg).}
	\label{tab:abolute_photometry}
	\centering
		\begin{tabular}{cccccccccc}
		\hline\hline
            $J_{\rm D}$  & Filter   &  $H$     &  $e_{\rm H}$          & $k$     & $e_{\rm k}$ & $\zeta$  & $r_{\rm H}$   & $\Delta$ & $\alpha$  \\
	                              &          & (mag)   & (mag)               &           &                   &             & (au)      & (au)         & ($^{\circ}$) \\
				       	\hline
       2456572.43011       & V       &  7.42    &  0.05  & 0.30     & 0.02          &  1.15   & 16.395 & 15.4756  & 1.4108 \\
       2456572.45304           & R       &  7.00     & 0.02   & 0.37     & 0.02         &    1.09    & 16.395 & 15.4756  & 1.4108  \\
    2457606.60204    &V  &  7.46     & 0.09                  &  0.09 & 0.05          & 1.68 - 1.14& 15.496 & 15.558 & 3.7375\\
    2457606.60204 & R      &  7.03     & 0.02                  &  0.06  & 0.04        & 1.65 - 1.16& 15.496 & 15.558 & 3.7375\\
       2457608.58729  &V&  7.49     & 0.08                  &  0.12 & 0.05          & 2.03 - 1.12& 15.494 & 15.545 & 3.7322\\
    2457609.60545  & V      &  7.46     & 0.03                  &  0.08 & 0.04          & 1.53 - 1.12& 15.494 & 15.508 & 3.7489\\
		\hline\hline
		\end{tabular}
\end{table*}

\section{Simple ellipsoid description}
\subsection{Pole determination (modeling of the light curve amplitude)}
\label{sec:aspect_angle}

\begin{table*}
   \caption[Light-curve amplitudes.]{Bienor's light-curve amplitudes from different epochs.}
\begin{threeparttable}
		\begin{tabular}{ccccccc}
		\hline\hline
		Date       & (year) &  $2001.626$          & $2004.781$           & $2014.930$                & $2015.951$ & $2016.597$\\
		\hline
	  $\Delta m$ & (mag)    &  $0.609 \pm 0.048^{1}$ & $0.34 \pm 0.08^{2}$  & $0.088 \pm 0.007^{\star}$ & $0.082 \pm 0.009^{\star}$& $0.10 \pm 0.02^{\star}$\\
		\hline\hline
		\end{tabular}
		\begin{tablenotes}
	\item $^{1}$ \cite{Ortiz2003a}, $^{2}$ \cite{Rabinowitz2007}, $^{\star}$ This work.
	\end{tablenotes}
	\end{threeparttable}
		\label{tab:amplitude_data}

\end{table*}

As can be seen in Table \ref{tab:amplitude_data}, the light curve amplitude has changed within the last 16 years from 0.609 mag \citep{Ortiz2002a}\footnote{We took the data published in that work to fit to a Fourier function as in section \ref{sec:light_curves} to determine our own $\Delta m$ value, which is slightly lower than that reported by \cite{Ortiz2002a}. This is because those authors took just the maximum and minimum of their data and subtracted them.} in 2000 to 0.082 mag in 2015 and actually it seems it is starting to increase slightly (this work, section \ref{sec:light_curves}). 
This implies that Bienor's aspect angle is evolving in time. We can take advantage of this to determine the orientation of the pole of the centaur as first done in \cite{Tegler2005} for centaur Pholus. We consider that Bienor is a Jacobi ellipsoid as in previous works \citep{Ortiz2002a, Rabinowitz2007}. Furthermore, as shown in the three rotational light curves (Figs. \ref{fig:light_curve_2014}, \ref{fig:light_curve_2015} and \ref{fig:light_curve_2016}) and in rotational light curves published in the aforementioned works, the maxima and minima of the Fourier function have different depths, which is another indication that the light curve is indeed mainly due to the body shape. The light curve amplitude produced by a triaxial body shape is given by the following equation:

\begin{equation}
\Delta m=-2.5\log\left(\frac{A_{\rm min}}{A_{\rm max}}\right),
\label{eq:lightcurve_amplitude}
\end{equation}
where $A_{\rm min}$ and $A_{\rm max}$ are the minimum and maximum area of the object given by:

\begin{equation}
\label{eq:area_min}
A_{\rm min}=\pi\,b\left[a^2\cos^2(\delta)+c^2\sin^2(\delta)\right]^{1/2}
\end{equation}
and
\begin{equation}
\label{eq:area_max}
A_{\rm max}=\pi\,a\left[b^2\cos^2(\delta)+c^2\sin^2(\delta)\right]^{1/2}
\end{equation}
where $a$, $b$ and $c$ are the semi-axes of the triaxial body (with $a>b>c$). Semi-axes ratios can be estimated under the assumption of hydrostatic equilibrium \citep{Chandrasekhar1987} and should comply with the fact that the effective diameter (in area) is $198^{+6}_{-7}$ km as determined by Herschel measurements \citep{Duffard2014}.  Finally, $\delta$ is the aspect angle, which is given by the ecliptic coordinates of the angular velocity vector (the pole direction) and the ecliptic coordinates of the object as follows:

\begin{equation}
\delta=\frac{\pi}{2}-\arcsin\left[\sin(\beta_{\rm e})\sin(\beta_{\rm p}) + \cos(\beta_{\rm e})\cos(\beta_{\rm p})\cos(\lambda_{\rm e}-\lambda_{\rm p})\right],
\label{eq:aspect_angle}
\end{equation}
where $\beta_{\rm e}$ and $\lambda_{\rm e}$ are the ecliptic latitude and longitude of the sub-Earth point in the Bienor-centred reference frame and $\beta_{\rm p}$ and $\lambda_{\rm p}$ are the ecliptic latitude and longitude of Bienor's pole \citep{Schroll1976}. 

We fitted the observational data from the literature and this work (see Table \ref{tab:amplitude_data}) to the Eq. (\ref{eq:lightcurve_amplitude}). We carried out a grid search for the quantities $\beta_{\rm p}$, $\lambda_{\rm p}$ and $b/a$ axis ratio, which gave theoretical values for $\Delta m$ with the smallest $\chi^2$ fit to the observed points. $\beta_{\rm p}$ and $\lambda_{\rm p}$ were explored on the entire sky at intervals of 5$^{\circ}$ and $b/a$ ratio was explored from 0.33 to 0.57 at steps of 0.04. The limiting values defining this interval, $0.33$ and 0.57, are chosen taking into account that relation between the $b/a$ and the light curve amplitude of the object. On the one hand, the upper limit $b/a=0.57$ is determined by the maximum light-curve amplitude observed for Bienor up to date \citep{Ortiz2003a}, namely the first point in Fig. \ref{fig:deltam_models} from year 2001. The measured amplitude implies that $b/a$ cannot be below $0.57$, as lower ratios would fail to provide such a rotational variability, independently of the value of the aspect angle. On the other hand, the lower limit gives $b/a=0.33$ arises from the condition that the light curve amplitude is always below $\Delta m=1.2$ mag. Light curve amplitudes which go above this value at some point in the evolution of the object are most likely caused by contact binary systems \citep{Leone1984}. 

In order to evaluate the goodness of the fit, we used a $\chi^2$ test as follows:

\begin{equation}
\label{eq:chi}
\chi^2_{\rm \Delta m} = \frac{\sum\left(\left(\Delta m_{\rm theo}-\Delta m_{\rm obs}\right)^2/e_{\rm \Delta m}^2\right)}{N_{\rm \Delta m}},
\end{equation}
where $\Delta m_{\rm theo}$ represents theoretical values, $\Delta m_{\rm obs}$ represents observational data, $e_{\rm \Delta m}$ represents errors for light curve amplitude observational data and $N_{\rm \Delta m}$ is the number of the light curve amplitude observational data.

The result was a pole solution of $\beta_{\rm p}=50\pm 3^{\circ}$ and $\lambda_{\rm p}=35\pm8^{\circ}$ and axes ratio of $b/a=0.45\pm0.05$ (see Table \ref{tab:pole_determination}). These values gave a $\chi^2_{\rm \Delta m}$ of 0.27 (the direction $\lambda_{\rm p}=215^{\circ}$ and $\beta_{\rm p}=-50^{\circ}$ is also possible for the same $\chi^2_{\rm \Delta m}$ value). In Fig. \ref{fig:deltam_models} the blue line represents the light curve amplitude model; also the observational data are shown. To estimate the uncertainties, we searched for all parameters within $\chi^2_{\rm \Delta m, min}$ and $\chi^2_{\rm \Delta m, min}+1$.

\begin{table*}
	\centering
   \caption{Results from the simplest modeling of the light curve amplitude (see Sec. \ref{sec:aspect_angle}). Columns are as follows: elongation ($b/a$); flattening ($c/b$), ecliptic latitude and longitude of Bienor's pole ($\beta_{\rm p}$, $\lambda_{\rm p}$), $\chi^2$ test from Eq. \ref{eq:chi} $(\chi^2_{\rm \Delta m})$.}
   \begin{threeparttable}
		\begin{tabular}{cccccccc}
		\hline\hline
	                       $b/a$ &    $c/b$             &	$\beta_{\rm p}$       & $\lambda_{\rm p}$ &  $\rho$                       &$\chi^2_{\rm \Delta m}$ & n & N \\
	                                 &                          &   	($^{\circ}$)              &  ($^{\circ}$)             &  (kg$\,$m$^3$)          &                                       &    &     \\
		\hline
                $0.45\pm0.05$ &  0.79$\pm$0.02 &  50$\pm$3                &  35$\pm$5              &     594$^{+47}_{-35}$ & 0.27                              & 3  & 5  \\
		\hline\hline
\end{tabular}
\begin{tablenotes} 
\item Note that the $\beta_{\rm p}=-50^{\circ}$ and $\lambda_{\rm p}=215^{\circ}$ solution is also valid.
\end{tablenotes}
\end{threeparttable}
		\label{tab:pole_determination}
\end{table*}

\subsection{Modeling of the absolute magnitude}
\label{sec:modelling_absolute_magnitude}

Right from the first observing runs that we carried out we realised that the amplitude of the rotational variability of Bienor had changed dramatically with respect to the first measurements in 2001, in which the amplitude was around 0.7 mag \citep{Ortiz2003a}. The usual explanation to that kind of changes in solar system bodies is related to a change in orientation of an elongated body. As explained in the previous section, using that approach for Bienor we came up with reasonable results. However, as discussed in the following this kind of model does not offer a satisfactory explanation of the large change in the absolute magnitude of Bienor in the last 16 years, which is considerably larger than what would be expected. In Table \ref{tab:absolute_magnitude_data} we show absolute magnitudes from the literature and from this work.
\begin{table*}
	\centering
		\caption{Bienor's absolute magnitude in different epochs. Abbreviations are as follows: absolute magnitude in V-band ($H_{\rm V}$), its associated error reported by the authors ($e_{\rm H_V}$), modified errors that we estimate as described in the Sec. \ref{sec:discussion} ($e'_{\rm H_V}$). The renormalization of the uncertainties bias the fit toward most recent values, as light-curve amplitudes are smaller in recent epochs.}
		\label{tab:absolute_magnitude_data}
		\begin{threeparttable}
		\begin{tabular}{ccccc}
		\hline\hline		
		Observation date       & $H_V$            & $e_{\rm H_V}$  &    $e'_{\rm H_V}$ & Bibliography \\
		                                  &  (mag)             &     (mag)            &      (mag)              &   \\
		\hline
		Nov. 2000                  &8.08                  &0.06                  &       0.31                & \cite{Delsanti2001} \\
	       May 2001         &    7.75             &    0.02                &      0.26                 &    \cite{Tegler2003}\\
		2002  $^{\dag}$        & 7.85$^{\star}$  &0.05                    & 0.26                      & \cite{Bauer2003} \\
	        Aug.  2002                 &  8.04               &0.02                    & 0.22                      & \cite{Doressoundiram2007} \\
		2004  $^{\ddag}$       & 7.588            &0.035                    & 0.035                    & \cite{Rabinowitz2007}\\
		Sep. 2007                  & 7.46              & 0.03                    & 0.13                       & \cite{DeMeo2009}\\ 
		Oct. 2013                  & 7.42                &   0.02                  & 0.04                      & This work \\
		Aug. 2016      &7.47                 &  0.04                    & 0.04                   & This work\\
		\hline\hline
				\end{tabular}
		\begin{tablenotes}
	Note: \cite{Doressoundiram2002} published another value of Bienor's magnitude which is in disagreement with the value for the same epoch published by T\cite{Tegler2003}. In order to check the correct value we searched for other values in the Minor Planet Center database for the same epoch. We found reliable data from  surveys in Johnson's R and V band that were in agreement with \cite{Tegler2003} but not with \cite{Doressoundiram2002}. For this reason we have not included the \cite{Doressoundiram2002} data point.
	\item  $^{\star}$ Value obtained from absolute magnitude in the $R$-band using the colour correction published by the authors.
	\item $^{\dag}$\citet{Bauer2003} observed Bienor on 29$^{th}$ October 2001 and on 13$^{th}$ June 2002. $^{\ddag}$ \citet{Rabinowitz2007} observed Bienor between July 2003 and December 2005.	
	\end{tablenotes}
	\end{threeparttable}
\end{table*}

The absolute magnitude of Bienor can be obtained using the previous values of the three parameters ($\beta_{p}$, $\lambda_{p}$ and $b/a$) by means of the following equation:

\begin{equation}
\label{eq:absolute_magnitude_noring}
H_{\rm V}=-V_{\odot}+2.5\log\left(\frac{C^2\pi}{p_{\rm V}\,A_{\rm B}(\delta)}\right),
\end{equation}
where $H_{\rm V}$ is the absolute magnitude of the object, $V_{\odot}=-26.74$ mag is the absolute magnitude of the Sun in V-band, $p_{\rm V}=0.043^{+0.016}_{-0.012}$ is the geometric albedo of the object in V-band \citep{Duffard2014}, $C = 1.496\times$10$^8$ km is a constant and $A_{\rm B}(\delta)$ is the rotational average area of the body in km$^2$, as determined from $A_{\rm min}$ and $A_{\rm max}$ given by Eqs. (\ref{eq:area_min}) and (\ref{eq:area_max}), respectively; with the constraint that the mean area matches the area for Bienor's effective diameter of 198$^{+6}_{-7}$ km determine by Herschel \citep{Duffard2014}.

This can be compared with the observational data shown in Table \ref{tab:absolute_magnitude_data}, as illustrated the blue line in Fig. \ref{fig:absolute_models}. It is apparent that the absolute magnitude observational data do not follow the curve obtained from Eq. (\ref{eq:absolute_magnitude_noring}). Indeed, we obtained a value of 195 for the $\chi^2$ test which now is defined as follows:

\begin{equation}
\chi^2_{\rm H_V} = \frac{\sum\left(\left(H_{\rm V,theo}-H_{\rm V,obs}\right)^2/e_{\rm H_V}^2\right)}{N_{\rm H_V}},
\label{eq:chi_H_V}
\end{equation}
where $H_{\rm V,theo}$ represents theoretical absolute magnitudes, $H_{\rm V,obs}$ represents observational data, $e_{\rm H_V}$ represents errors for absolute magnitude observational data and $N_{\rm H_V}$ is the number of the absolute magnitude observational data.
\section{More complex models}
\subsection{Simultaneous modeling of absolute magnitude and light curve amplitude ({\it HE model})}
\label{sec:simultaneous_modelling}

Given this situation, one might wonder whether it would be possible to find a set of values for the parameters $\beta_{\rm p}$, $\lambda_{\rm p}$ and $b/a$ axis ratio, leading to a satisfactory fit for both Eqs. (\ref{eq:lightcurve_amplitude}) and (\ref{eq:absolute_magnitude_noring}) simultaneously. To check the viability of this possibility, we defined a $\chi^2_{\rm T}$ value so as to evaluate both fits at the same time as follows:

\begin{equation}
\label{eq:chi_pdf}
\chi^2_{\rm T}=\frac{\left(\chi^2_{\rm \Delta m}+\chi^2_{\rm H_V}\right)}{2}.
\end{equation}
Here $\chi^2_{\rm \Delta m}$ is the $\chi^2$ value from Eq. (\ref{eq:chi}) and $\chi^2_{\rm H}$ is the $\chi^2$ value from Eq. (\ref{eq:chi_H_V}). The $\chi^2_{\rm T}$ value obtained using the original values of the parameters ($\beta_{\rm p}$, $\lambda_{\rm p}$ and $b/a$) as in the previous section was $\sim$ 98. We searched for other sets of values that could fit satisfactorily both equations; nevertheless, all possible parameters gave poor fits with $\chi^2_{\rm T}>45$.

Hence, we did not find any solution that can fit satisfactorily the observational data of the absolute magnitude. As shown in Fig. \ref{fig:absolute_models}, there is an increase of brightness with time that is not explained by the {\it hydrostatic model}. This leads us to think that there must be some physical process which produces such a large slope in the observational data, which we are not taking into account. This physical process might have to do with the existence of material orbiting around Bienor with a ring shape. The reflected flux contribution due a ring plus the reflected flux contribution due to the body, as in the cases of Chariklo and Chiron \citep{Braga-Ribas2014,Ortiz2015}, could produce the strong drop which is shown in the observational data of the absolute magnitude (see Sec. \ref{sec:ring_system}). However, other scenarios might be also possible. In the following we discuss the different scenarios that we have considered.

\subsection{Modeling of the data relaxing both the albedo and the effective diameter constrains from Herschel ({\it Herschel} model)}
\label{sec:relaxing_albedo_diameter}

If one takes a look at the absolute magnitude plot it seems like the model in blue in Fig. \ref{fig:absolute_models} is displaced up with respect to the data points. This means that the albedo or the effective diameter could be higher than we used for the model, or even a combination of both of them. Therefore, we search around the values given by \cite{Duffard2014} taking into account their error bars. As a result the fit was improved using an albedo of 5.7\% and an effective diameter of 204 km (see yellow line in Fig. \ref{fig:absolute_models}). This last fit modifies slightly the pole direction obtained in Sec. \ref{sec:aspect_angle} (see Table \ref{tab:models}, this model is referred to as Herschel). However, the model is still poor and does not represent the drop of the observational points.

\subsection{Modeling of the data relaxing the assumption of hydrostatic equilibrium ({\it NHE} model)}
\label{sec:no_equilibrium}

We thought about the possibility that Bienor might not be in perfect hydrostatic equilibrium, this is possible because Bienor's size is small enough to allow for departures of hydrostatic equilibrium. Therefore, we also searched the aforementioned grid adding the $c/b$ axis ratio from 0.5 to 1.0 at intervals of 0.1. The smallest $\chi^2_{\rm T}$, which was equal to 13.63, provides us the following parameters: 60$^{\circ}$, 25$^{\circ}$, 0.33 and 0.5 for $\beta_{\rm p}$, $\lambda_{\rm p}$, $b/a$ and $c/b$, respectively. However, theses ratios imply an extremely elongated body with an a-axis around 6 times bigger than the c-axis. There is no known body in the Solar system with this extremely elongated shape for such a large body as Bienor. Hence we do not think that this is plausible. This model is referred to as {\it NHE} in Table \ref{tab:models}.

However, we tried to find a good fit simultaneously relaxing the Herschel constrains as in the last subsection. This search provides a possible solution for an albedo of 5.1\% and using the effective diameter given by Herschel with a pole direction of 50$^{\circ}$, 30$^{\circ}$ for $\beta_{\rm p}$, $\lambda_{\rm p}$, respectively and axis ratio of 0.33 and 0.7 for $b/a$ and $c/b$, respectively. This model is plotted in Figs. \ref{fig:deltam_models} and \ref{fig:absolute_models} (see orange line) and referred as {\it NHE- Herschel} model in Table \ref{tab:models}.

\subsection{Modeling of the data with the inclusion of variable geometric albedo ({\it Albedo} model)}
\label{sec:variable_albedo}

\begin{figure}
	\includegraphics[width=\columnwidth]{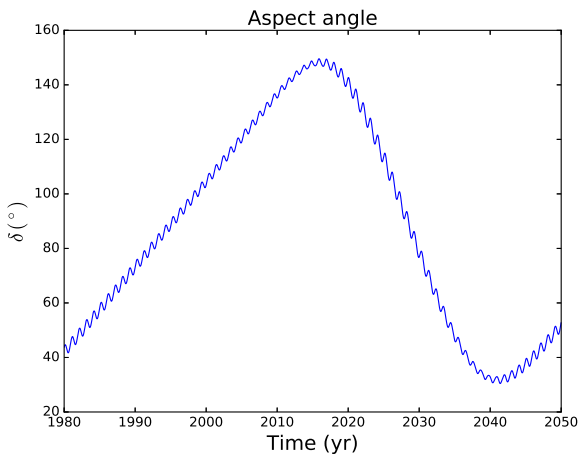}
    \caption{Bienor's aspect angle versus time. The black line shows the result of the equation (\ref{eq:aspect_angle}) with $\lambda_p=35^{\circ}$ and $\beta_p=50^{\circ}$. The edge-on position (when the angle between the spin axis orientation and the line of sight is 90$^{\circ}$) is achieved around 1988 and around 2030.}
    \label{fig:aspectangle_eq_noring}
\end{figure}

Another possibility to increase the brightness of Bienor beyond the values of the modeling in Sec. \ref{sec:modelling_absolute_magnitude} is to increase the geometric albedo of the body as a function of time or as a function of aspect angle. If the polar regions of Bienor have very high albedo, then it might be possible that Bienor becomes brighter as seen from Earth simply because we observe higher latitudes of Bienor as the aspect angles changes (because the current aspect angle in 2015 is $\sim 150^{\circ}$, see Fig. \ref{fig:aspectangle_eq_noring}). The needed change of Bienor's geometric albedo is from 3.9\% in 2000 to 7.6\% in 2008, when the aspect angle is ~100$^{\circ}$ and ~130$^{\circ}$, respectively. This is shown in Fig. \ref{fig:absolute_models} (green line). Following this situation, one can extrapolate the albedo which the object would have if its aspect angle is 180$^{\circ}$. Under this assumption, the albedo should be around 10\%. Such a dramatic change in Bienor's geometric albedo is hard to explain because the polar caps  would have to have an even higher albedo than 10\% (which is the hemispheric average). For instance, a polar cap covering around 42\% of the total area (viewed from the top) with an albedo of 16\% would fit the data. A more confined polar cap would have to have an even higher albedo than 16\%. It is difficult to find a mechanism that would cause such a large north-south asymmetry on the surface. For reference, the maximum longitudinal albedo variability on Bienor is only around a few percent because the two maxima of the rotational light curve differ by only 0.05 mag. We refer to this model as ({\it Albedo} model) in Table \ref{tab:models}.

\subsection{Modeling of the data with the inclusion of a ring system ({\it Ring} model)}
\label{sec:ring_system}

In order to explain the observational data of the absolute magnitude, we included a ring contribution in the aforementioned equations: (\ref{eq:lightcurve_amplitude}) and (\ref{eq:absolute_magnitude_noring}), see Sec. \ref{eq:aspect_angle}. On the one hand, the light curve amplitude produced by the system Bienor + ring, $\Delta m_{\rm S}$, is now given as follows:

\begin{equation}
\label{eq:lightcurve_amplitude_ring}
\Delta m_{\rm S}=-2.5\log\left(\frac{A_{\rm min}\,p_{\rm B} +A_{\rm R}\,p_{\rm R}}{A_{\rm max}\,p_{\rm B}+A_{\rm R}\,p_{\rm R}}\right).
\end{equation}

The additional parameters are the ring's area ($A_{\rm R}$), the ring's albedo ($p_{\rm R}$) and Bienor's albedo ($p_{\rm B}$). $A_{\rm R}$ is given by

\begin{equation}
A_{\rm R}=\pi\mu\left( (R_{\rm R}+d_{\rm R})^2-R_{\rm R}^2\right),
\end{equation}
where $\mu=|\cos(\delta)|$ ($\delta$ is the aspect angle, see Sec. \ref{eq:aspect_angle}); $R_{\rm R}$ is the ring's radius and $d_{\rm R}$ is the ring's width. On the other hand, the absolute magnitude of the system, $H_{\rm S}$, is now given as follows:

\begin{equation}
H_S= -V_{\odot} + 2.5\log\left(\frac{C^2\pi}{A_{\rm B}\,p_{\rm B}+A_{\rm R}\,p_{\rm R}}\right).
\label{eq:absolute_magnitude_ring}
\end{equation}

\begin{figure*}
	\includegraphics[width=0.99\textwidth]{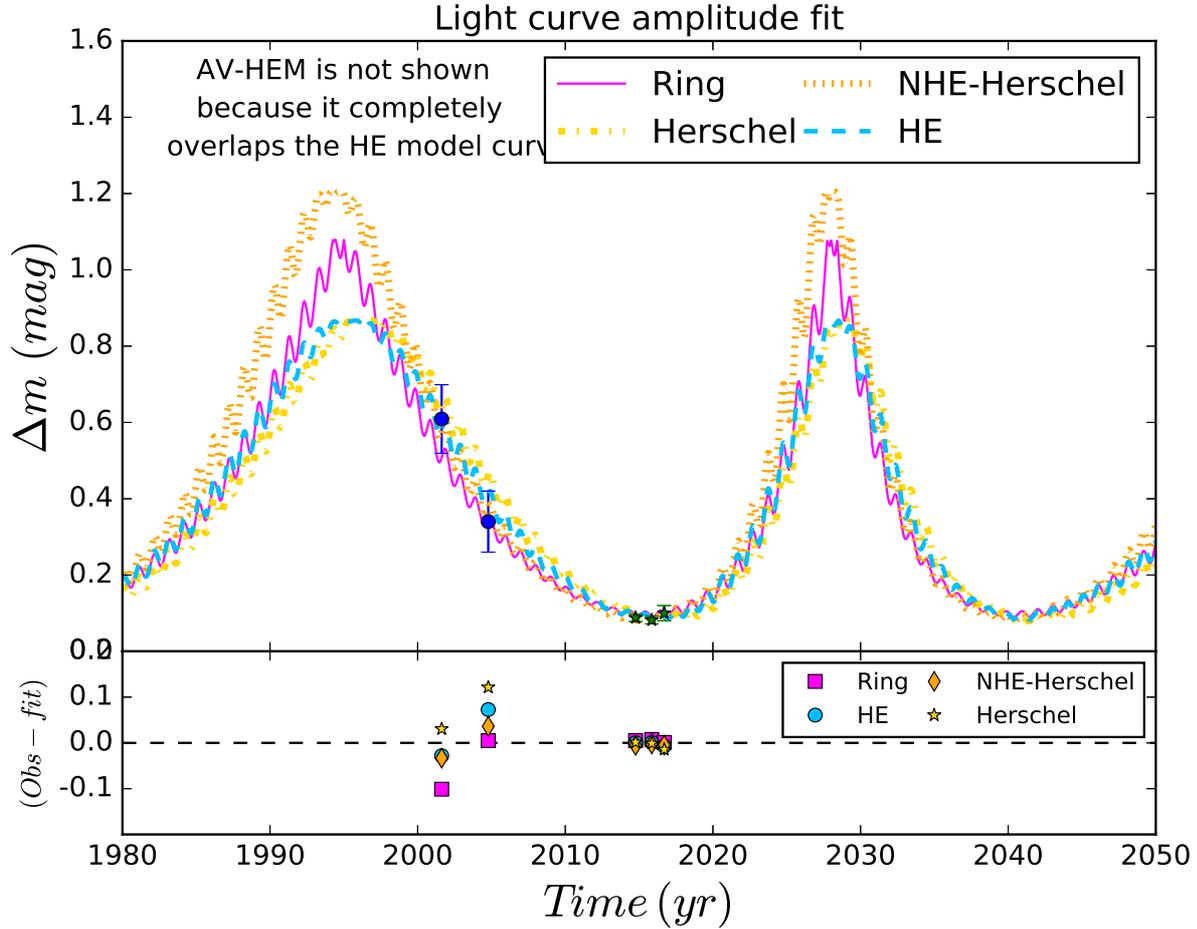}
    \caption{Bienor's light curve amplitude fit. At the top panel: The blue dashed line represents the hydrostatic equilibrium model ({\it HE} model, see Sec. \ref{sec:aspect_angle}). The pink line represents the ring system model ({\it Ring} model, see Sec. \ref{sec:ring_system}). The yellow dotted line represents the hydrostatic equilibrium model relaxing Herschel constrains ({\it Herschel} model, see Sec. \ref{sec:relaxing_albedo_diameter}). The orange dotted line represents the no hydrostatic equilibrium model relaxing Herschel constrains ({\it NHE-Herschel} model, see Sec. \ref{sec:no_equilibrium}). Dark blue circle points show data taken from literature. Green star points show data from this work. Bottom panel: residuals of the observational data with respect to the different models. Blue circle points correspond to the hydrostatic equilibrium model. Pink square points correspond to the ring system model. Yellow star points correspond to the hydrostatic equilibrium model relaxing Herschel constrains. Orange diamond points correspond to the no hydrostatic equilibrium model relaxing Herschel constrains. {\it Albedo} model is not shown because it completely overlaps the {\it HE} model curve.}    \label{fig:deltam_models}
\end{figure*}

\begin{figure*}
	\includegraphics[width=0.99\textwidth]{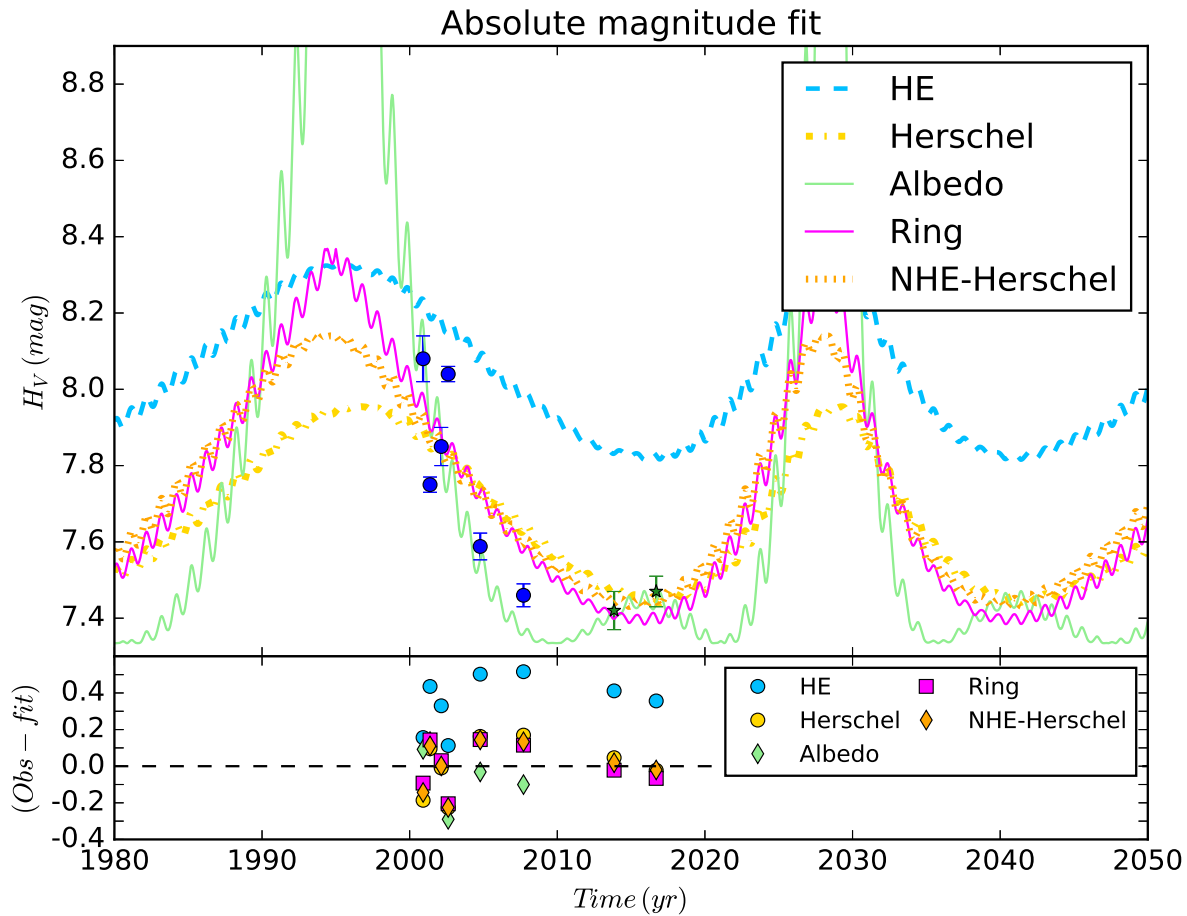}
    \caption{Bienor's absolute magnitude fit. The blue dashed line represents the hydrostatic equilibrium model ({\it HE} model, see Sec. \ref{sec:aspect_angle}). The pink line represents the ring system model ({\it Ring} model, see Sec. \ref{sec:ring_system}). The yellow dotted line represents the hydrostatic equilibrium model relaxing Herschel constrains ({\it Herschel} model, see Sec. \ref{sec:relaxing_albedo_diameter}). The orange dotted line represents the no hydrostatic equilibrium model relaxing Herschel constrains ({\it NHE-Herschel} model, see Sec. \ref{sec:no_equilibrium}). The green line represents the albedo variability model ({\it Albedo} model, see Sec. \ref{sec:variable_albedo}). The dark blue circle points show data taken from literature with errors reported by the authors. Green star points show data from this work. At the bottom panel: residual of the observational data. Blue star points correspond to the hydrostatic equilibrium model. Pink square points correspond to the ring system model. Yellow circle points correspond to hydrostatic equilibrium model relaxing Herschel constrains. Orange diamond points correspond to no hydrostatic equilibrium model relaxing Herschel constrains. Green diamond points represent albedo variability model.}
    \label{fig:absolute_models}
\end{figure*}

The same exercise as in Sec. \ref{sec:aspect_angle} was carried out. We explored the quantities $\lambda_{\rm p}$, $\beta_{\rm p}$, $a/b$, $p_{\rm B}$, $p_{\rm R}$, $A_{\rm R}$ and $R_{\rm eff}$ (Bienor's effective radius) in equations (\ref{eq:lightcurve_amplitude_ring}) and (\ref{eq:absolute_magnitude_ring}) that gave theoretical values for both fits, light curve amplitude and absolute magnitude, which minimize the difference between observational and theoretical data. $A_{\rm R}$ was explored from 4000 km$^2$ to 10000 km$^2$ at intervals of 500 km$^2$, the effective radius from 90 km to 99 km at intervals of 3 km. We also explored ring's albedo from 8\% to 16\% at steps of 2\% and Bienor's albedo from 3\% to 6\% at steps of 1\%. We should take into account that the solution of this problem is degenerated as different ring sizes combined with different nucleus sizes could fit to the data points. Here, we only want to note that a ring solution is plausible. A hypothetical ring of around 10000 km$^2$ is around 2 times smaller than Chiron's ring as shown in \cite{Ortiz2015} and 1.2 times larger than that of Chariklo. A dense and narrow ring of 315 km inner radius and 318 km outer radius would do the job with no modification of the pole direction obtained in Sec. \ref{sec:aspect_angle}. The best fit for observational data provides the following values: $a/b=0.37\pm0.10$, $p_{\rm B}=5.0\pm0.3$\%, $p_{\rm R}=12.0\pm1.5$\%, $A_{\rm R}=6000\pm700$ and $D_{\rm eff}=180\pm5$ km (see the pink lines in Figs. \ref{fig:deltam_models} and \ref{fig:absolute_models}) This is the best model in terms of $\chi^2_{\rm T}$. We refer to this model as ({\it Ring} model) in Table \ref{tab:models}.


\begin{table*}
	\centering
   \caption{Bienor's parameters for each model to simultaneously fit light curve amplitudes and absolute magnitudes using errors reported by the authors. The columns contain the following information: Model designation (see foot note); elongation ($b/a$); flattening ($c/b$); ecliptic latitude and longitude of Bienor's pole ($\lambda_{\rm p}$, $\beta_{\rm p}$); Bienor's albedo in V-band ($p_{\rm B}$); Bienor's effective diameter ($D_{\rm eff}$); ring's area ($A_{\rm R}$); ring's albedo in V-band ($p_{\rm R}$); goodness of the fit given by the Eq. \ref{eq:chi} ($\chi_{\rm \Delta m}$); goodness of the fit given by the Eq. \ref{eq:chi_H_V} ($\chi_{\rm H}$); goodness of the simultaneous fit to the Eqs. \ref{eq:chi} and \ref{eq:chi_H_V}  ($\chi^2_{\rm T}$); number of parameters of the fit (n); number of light curve amplitude and absolute magnitude data ($N_{\rm \Delta m}$, $N_{\rm H}$).}
		\begin{threeparttable}
		\begin{tabularx}{\textwidth}{p{1.0cm}cccp{1.0cm}cXp{1.0cm}p{0.8cm}p{0.8cm}Xp{0.25cm}p{0.25cm}p{0.1cm}p{0.2cm}p{0.2cm}}
		\hline\hline
		Model                        &$b/a$                 &$c/b$                     &$\beta_{\rm p}$               &$\lambda_{\rm p}$                 &$p_{\rm B}$                     &$D_{\rm eff}$                               &$A_{\rm R}$       &$p_{\rm R}$          & $\rho$                     &$\chi^2_{\rm T}$     & $\chi^2_{\rm \Delta m}$       &  $\chi^2_{\rm H}$                  & n                                 &$N_{\rm \Delta m}$        & $N_{\rm H}$\\
                        &        &        &     ($^{\circ}$)        &     ($^{\circ}$)  & (\%)          & (km)  & (km$^2$)     &  (\%)            &    (kg$\,$m$^{-3}$) &  &   &  &&\\
		\hline
{\it HE}                                            &0.33$\pm$0.02   &0.85$\pm$0.01     &25$\pm$7                        &15$\pm$6                              & 4.3$^{+1.2}_{-1.6}$        &198$^{+6}_{-7}$    &                           &                             &   $742^{+41}_{-35}$ &     45.3                                   & 18.8                              &      71.8          & 3                                  & 5                                   & 7 \\
 {\it Herschel}                                          &0.45$\pm$0.08   &0.79$\pm$0.04     &50$\pm$4                         &40$\pm$7               & 5.7 $\pm$0.2                                 &204$\pm$4        &                           &                             & $594^{+84}_{-52}$   &    14.08                                       & 0.66                              &      27.5         & 5                                 & 5                                    & 7\\
{\it NHE}                                              &0.33$\pm$0.03     &0.5$\pm$0.04          &  60 $\pm$5                     & 25$\pm$5                 & 4.3 $^{+1.2}_{-1.6}$                 & 198$^{+6}_{-7}$ &                          &                                &                                &  13.63                                          & 1.52                             &      25.7         & 4                                  & 5                                    & 7\\
{\it  NHE-Herschel}                                     &0.33$\pm$0.05     &0.7$\pm$0.05       & 50$\pm$3                      &  30$\pm$6                & 5.1$\pm$0.2                            & 198$\pm$3 &                          &                               &                                 &  12.69                                           & 0.23                              &      25.2         & 6                                 & 5                                      & 7\\
{\it Albedo}                                               &0.45$\pm$0.09               &0.79$\pm$0.04       &50$\pm$3                       & 35$\pm$3                  & 3.9 - 7.6                                  & 198$^{+6}_{-7}$ &                          &                               & $594^{+98}_{-57}$  &   17.07                                          & 0.27                              &      33.9         & 4                                  & 5                                     & 7\\
{\it Ring}                                              &  0.37$\pm$0.10              &0.83$\pm$0.06       &55$\pm$3                  & 30$\pm$5                      & 5.0$\pm$0.3                            & 180$\pm$5     & 6000$\pm$700& 12.0$\pm$1.5             &  $678^{+209}_{-100}$     &   12.40                                              & 0.40                              &      24.4         &7                                 &  5                                     & 7\\
	         		\hline\hline
		\end{tabularx}
		\begin{tablenotes}
		\item {\it HE}: Hydrostatic Equilibrium Model\\
		\item {\it Herschel} : Hydrostatic Equilibrium Model relaxing Herschel constrains (Sec. \ref{sec:relaxing_albedo_diameter}).\\
		\item {\it NHE}: No Hydrostatic equilibrium Mode (Sec. \ref{sec:no_equilibrium}).\\
		\item {\it NHE-Herschel}: No Hydrostatic Equilibrium Model relaxing Herschel constrains (Sec. \ref{sec:no_equilibrium}).\\
		\item {\it Albedo}: Albedo Variability Model (Sec. \ref{sec:variable_albedo}).\\
		\item {\it Ring}: Ring System Model (Sec. \ref{sec:ring_system}).
				\end{tablenotes}
	\end{threeparttable}
		\label{tab:models}
\end{table*}

\section{Discussion}
\label{sec:discussion}

We have considered several scenarios to simultaneously explain Bienor's change of light curve amplitude and absolute magnitude in the last 16 years. As can be seen in Fig. \ref{fig:absolute_models} at least two of the models seem to fit the data qualitatively, but the too high values of the goodness of fit test (Table \ref{tab:models}) indicate that either the models are not fully satisfactory or that the errors have been underestimated. Indeed, we have reasons to suspect that the absolute magnitudes determined by several authors could have been affected by the large rotational variability of Bienor in those years. Hence, we revised the errors with the very conservative strategy of assigning an extra uncertainty of half the full amplitude of the rotational light curve at each epoch.

When the revised errors in Table \ref{tab:absolute_magnitude_data} are used for the computation of new values of the goodness of fit, slightly different fits with respect to Table \ref{tab:models} are obtained. They are summarized in Table \ref{tab:models_big_errors}. Now the goodness of fit test provides too low values for some models, possibly indicating that the errors have been overestimated in this case. Given that an accurate determination of errors in the absolute magnitudes was not possible, we suspect that the reality probably falls in between the two different error estimates and therefore the best model fits should be something in between the results of Table \ref{tab:models} and Table \ref{tab:models_big_errors}.

As can be seen in aforementioned tables {\it HE} model gives far poorer fits than the other models; therefore, we can conclude that a simple hydrostatic equilibrium model cannot fit the data.  By relaxing the albedo and effective diameter constrains given by Herschel \citep{Duffard2014} we improved the fit. A better solution is found by relaxing both the assumption of hydrostatic equilibrium and Herschel constraints. But concerning this model, the main difficulty is that it requires a very extreme body with too large $a/c$ ratio to be realistic for bodies of Bienor's size. Nevertheless, there are models of dumb-bell shaped contact binaries that can give rise to $a/c$ axial ratios of up to 4.14 \citep{Descamps2015}. Such a contact binary would not perfectly fit the data but together with a north-south asymmetry in the albedo might be close to offer a good solution. Using the formalism in \cite{Descamps2015} the $a/c=4.14$ axial ratio (approximately the axial ratio obtained in Sec. \ref{sec:no_equilibrium} when the Herschel constrains are relaxed) would require a density of 970 kg$\,$m$^{-3}$ for Bienor, given its known 9.1713 h period. Such a density in TNOs is expected for objects with an effective diameter around 500 km \cite[see supplementary material in][]{Ortiz2012b,Carry2012}; such a diameter is 2.5 times bigger than Bienor's effective diameter. Nevertheless, 970 kg$\,$m$^{-3}$ can not be completely discarded.

Concerning the albedo variability model, this would require a very bright polar cap on Bienor whereas the equatorial parts would have a geometric albedo of only a few percent. No centaur or small TNO has ever been found to exhibit such a remarkable albedo variability in its terrain. Polar caps of ices may be expected in objects with evaporation and condensation cycles, which does not seem to be viable for centaurs, because CO$_2$ would be too volatile and H$_2$O is not sufficiently volatile with the temperatures involved at the distances to the Sun in which Bienor moves. Hence, even though this is a possibility, it does not seem very promising.

For all of the above we thought about the possibility that Bienor could have a ring system, or a partial ring system because we know at least two other similar sized centaurs that have ring material around them \citep{Braga-Ribas2014,Ortiz2015}. When this possibility was considered we got a slightly better solution than for the no hydrostatic equilibrium model relaxing the Herschel constrains, with no modification of the pole direction that was obtained from the light curve amplitude fit in the case in which no ring is included (see Sec. \ref{sec:aspect_angle}).

On the other hand, we know that there is water ice detection in Bienor's spectra already reported in the literature \cite[e. g.][]{Dotto2003,Barkume2008,Guilbert2009}, which would also be consistent with the idea that Bienor could have an icy ring or icy ring material around its nucleus. This has been the case for centaurs Chariklo and Chiron, which also have spectroscopic detection of water ice and the variation of the depth of the ice features in their spectra is well explained due to a change in the aspect angle of the rings. This was a clear indication that the water ice is in the rings of these centaurs \citep{Duffard2014b,Ortiz2015}. Hence, the presence of water ice in the spectrum of centaur Bienor is also a possible indication of a ring around Bienor's nucleus. In fact, all centaurs that exhibit a water ice feature in their spectrum may be suspect of having a ring system.

Besides, the density we derive for Bienor with the model that includes a ring system (742 kg$\,$m$^{-3}$ in Table \ref{tab:models_big_errors}, 678 kg$\,$m$^{-3}$ in Table \ref{tab:models}) is slightly higher than what we derive without the inclusion of a ring system (594 kg$\,$m$^{-3}$, see Table \ref{tab:models}). The higher value looks somewhat more realistic because we already know (with high accuracy) the density of comet 67P from the Rosetta visit \cite[533 $\pm$ 0.006 kg$\,$m$^{-3}$ according to ][]{Patzold2016}. It would be somewhat surprising that the density of Bienor, which is much larger than comet 67P would be nearly identical, as we expect less porosity for larger bodies \cite[e. g.][]{Carry2012}.

Therefore the model with a ring not only explains the photometry, but it also results in a density value that seems more realistic. Hence a putative ring offers a more consistent physical picture than a huge albedo North-South asymmetry in the surface of Bienor or the other models, although combinations of the three different scenarios discussed may also give a satisfactory fit to the data. Hence, even though we favor the possibility that Bienor could have ring system, it is not firmly proven.

Future stellar occultations by Bienor may ultimately confirm or reject the existence of a dense ring system. In this regard, there will be two potentially good stellar occultation by Bienor on February 13$^{th}$ and December 29$^{th}$ 2017. These are occultations of bright enough stars so that detection of ring features is feasible and occur in highly populated areas of the world. Observations of these events and other future stellar occultations by Bienor, as well as spectroscopic observations, will indeed be valuable.

It must be noted that the derivation of the spin axis direction is not highly dependent on the different models so we have derived a relatively well constrained spin axis direction of $\lambda_{\rm p}=25^{\circ}$ to $40^{\circ}$ and $\beta_{\rm p}=45^{\circ}$ to $55^{\circ}$. Note that the symmetric solution $\lambda_{\rm p}=25^{\circ}+180^{\circ}$ to $40^{\circ}+180^{\circ}$ and $\beta_{\rm p}=-45^{\circ}$ to $-55^{\circ}$ is also possible. Besides, despite the different models we have derived a well constrained density between 550 and 1150 kg$\,$m$^{-3}$ in the most extreme cases.

\begin{table*}
	\centering
   \caption[Models.]{Bienor's parameters for each model to simultaneously fit light curve amplitudes and absolute magnitudes using errors taking into account the light curve amplitude. The columns contain the following information: Model designation (see foot note); elongation ($b/a$); flattening ($c/b$); ecliptic latitude and longitude of Bienor's pole ($\lambda_{\rm p}$, $\beta_{\rm p}$); Bienor's albedo in V-band ($p_{\rm B}$); Bienor's effective diameter ($D_{\rm eff}$); ring's area ($A_{\rm R}$); ring's albedo in V-band ($p_{\rm R}$); goodness of the fit given by the Eq. \ref{eq:chi} ($\chi_{\rm \Delta m}$); goodness of the fit given by the Eq. \ref{eq:chi_H_V} ($\chi_{\rm H}$); goodness of the simultaneous fit to the Eqs. \ref{eq:chi} and \ref{eq:chi_H_V} ($\chi^2_{\rm T}$); number of parameters of the fit (n); number of light curve amplitude and absolute magnitude data ($N_{\rm \Delta m}$, $N_{\rm H}$).}
		\begin{threeparttable}
		\begin{tabularx}{\textwidth}{p{1.0cm}cccp{1.0cm}cXp{1.0cm}p{0.8cm}p{0.8cm}Xp{0.25cm}p{0.25cm}p{0.1cm}p{0.2cm}p{0.2cm}}
		\hline\hline
		Model                        &$b/a$                 &$c/b$                     &$\beta_{\rm p}$               &$\lambda_{\rm p}$                 &$p_{\rm B}$                     &$D_{\rm eff}$                               &$A_{\rm R}$       &$p_{\rm R}$          & $\rho$                     &$\chi^2_{\rm T}$                         & $\chi^2_{\rm \Delta m}$      &  $\chi^2_{\rm H}$ & n                                 &$N_{\rm \Delta m}$        & $N_{\rm H}$\\
                &     &        &         ($^{\circ}$)        &     ($^{\circ}$)  & (\%)          & (km)  & (km$^2$)     &  (\%)            &    (kg$\,$m$^{-3}$) & & &   & && \\
		\hline
{\it HE}                                            &0.33$\pm$0.03   &0.85$\pm$0.02     &40$\pm$5                       &20$\pm$9                              & 4.3$^{+1.2}_{-1.6}$        &198$^{+6}_{-7}$    &                           &                             &   $742^{+64}_{-51}$ &    16.4                                      & 4.18                              &      28.7   & 3                                  & 5                                   & 7 \\
{\it Herschel}                          &0.45$\pm$0.07   &0.79$\pm$0.03     &50$\pm$3                         &35$\pm$11               & 5.9 $\pm$0.6                                 &204$\pm$10        &                           &                             & $594^{+71}_{-47}$   &    0.89                                      &  0.27                              &      1.52   &5                                 & 5                                    & 7\\
{\it NHE}                                              &0.33$\pm$0.05     &0.5$\pm$0.04          &  60 $\pm$3                     & 25$\pm$8                 & 4.3 $^{+1.2}_{-1.6}$                 & 198$^{+6}_{-7}$ &                          &                                &                                & 2.53                                          & 1.52                              &      3.54   & 4                                  & 5                                    & 7\\
{\it  NHE-Herschel}                             &0.33$\pm$0.08     &0.85$\pm$0.07       & 45$\pm$3                      &  30$\pm$7                & 5.9$\pm$0.5                            & 200$\pm$8            &                          &                               &                                 &  0.58                                           & 0.28                              &      0.89   & 6                                 & 5                                      & 7\\
{\it Albedo}                                               &0.45$\pm$0.07               &0.79$\pm$0.03       &50$\pm$3                       & 35$\pm$3                  & 3.9-7.6                                  & 198$^{+6}_{-7}$ &                          &                               & $594^{+71}_{-47}$  &   0.34                                          & 0.27                              &      0.51   & 4                                  & 5                                     & 7\\
{\it Ring}                                              &  0.33$\pm$0.12              &0.85$\pm$0.06       &50$\pm$5                  & 25$\pm$10                      & 5.0$\pm$0.5                            & 192$^{+10}_{-10}$     & 4000$\pm$1300& 12.0$\pm$0.4             &  $742^{+401}_{-149}$     &                     0.63 &      1.00   &0.81                                              &7                                 &  5                                     & 7\\
	         		\hline\hline
		\end{tabularx}
		\begin{tablenotes}
		\item {\it HE}: Hydrostatic Equilibrium Model\\
		\item {\it Herschel} : Hydrostatic Equilibrium Model relaxing Herschel constrains (Sec. \ref{sec:relaxing_albedo_diameter}).\\
		\item {\it NHE}: No Hydrostatic equilibrium Mode (Sec. \ref{sec:no_equilibrium}).\\
		\item {\it NHE-Herschel}: No Hydrostatic Equilibrium Model relaxing Herschel constrains (Sec. \ref{sec:no_equilibrium}).\\
		\item {\it Albedo}: Albedo Variability Model (Sec. \ref{sec:variable_albedo}).\\
		\item {\it Ring}: Ring System Model (Sec. \ref{sec:ring_system}).\\
		\item {The $\chi^2_{\rm T}$ values were obtained with the revised errors $e'_{\rm H_V}$ of Table \ref{tab:absolute_magnitude_data}.}
		\end{tablenotes}
	\end{threeparttable}
		\label{tab:models_big_errors}
\end{table*}

\section{Conclusions}
\label{sec:conclusions}

Thanks to several photometry runs in which we observed a remarkable change in the amplitude of the rotational variability of Bienor since 2000, and together with data available in the literature, we have been able to determine the orientation of the pole of Bienor ($\beta_{\rm p}=50^{\circ}$, $\lambda_{\rm p}=30^{\circ}$), and we have derived its shape ($b/a=0.45$). These results, together with the known rotation period allowed us to determine a density for Bienor of 594 kg$\,$m$^{-3}$ under the usual assumption of hydrostatic equilibrium. However, we find that the absolute magnitude of Bienor observed in different epochs is not compatible with a simple triaxial ellipsoid shape. We have investigated several possible scenarios to explain the anomalous absolute magnitude decline. We find that the inclusion of a thin ring system can explain the observed variation although other scenarios cannot be discarded. The required ring system's albedo and width are similar to those found in Chariklo and Chiron. When the ring system is included, the shape of Bienor's nucleus has to be somewhat more elongated and the resulting density is in between 688 and 742 kg$\,$m$^{-3}$, slightly higher than in the case in which no ring is considered. Future stellar occultation may shed light on the possible existence of a ring. To put the results in context, density, shape and pole orientation are important physical parameters that have been determined for only 3 other centaurs.

\section*{Acknowledgements}

We are grateful to the NOT, CAHA and OSN staffs. This research is partially based on observations collected at Centro Astron\'omico Hispano Alem\'an (CAHA) at Calar Alto, operated jointly by the Max-Planck Institut fur Astronomie and the Instituto de Astrof\'isica de Andaluc\'ia (CSIC). This research was also partially based on observation carried out at the Observatorio de Sierra Nevada (OSN) operated by Instituto de Astrof\'isica de Andaluc\'ia (CSIC). This article is also based on observations made with the Nordic Optical Telescope, operated by the Nordic Optical Telescope Scientific Association at the Observatorio del Roque de los Muchachos, La Palma, Spain, of the Instituto de Astrof\'isica de Canarias. Funding from Spanish grant AYA-2014-56637-C2-1-P is acknowledged, as is the Proyecto de Excelencia de la Junta de Andaluc\'ia, J. A. 2012-FQM1776. R.D. acknowledges the support of MINECO for his Ramon y Cajal Contract. The research leading to these results has received funding from the European Union's Horizon 2020 Research and Innovation Programme, under Grant Agreement no 687378. We thank the referee Dr. Benoit Carry for very helpful comments.







%
%


\bsp	
\label{lastpage}
\end{document}